# Energy Localization in Spherical Non-Hermitian Topolectrical Circuits

Xizhou Shen,[1] Xiumei Wang,[2*] Haotian Guo,[3] and Xingping Zhou[4*]

[1] *College of Integrated Circuit Science and Engineering, Nanjing University of Posts and Telecommunications, Nanjing 210003, China*

[2] *College of Electronic and Optical Engineering, Nanjing University of Posts and Telecommunications, Nanjing 210003, China*

[3] *School of Design Arts and Media Nanjing University of Science and Technology, Nanjing, China*

[4] *Institute of Quantum Information and Technology, Key Lab of Broadband Wireless Communication and Sensor Network Technology, Ministry of Education, Nanjing University of Posts and Telecommunications, Nanjing 210003, China*

*\*wxm@njupt.edu.cn*

*\*zxp@njupt.edu.cn*

This work delves into the energy localization in non-Hermitian systems, particularly focusing on the effects of topological defects in spherical models. We analyze the mode distribution changes in non-Hermitian Su-Schrieffer-Heeger (SSH) chains impacted by defects, utilizing the Maximum Skin Corner Weight (MaxWSC). By introducing an innovative spherical model, conceptualized through bisecting spheres into one-dimensional chain structures, we investigate the non-Hermitian skin effect (NHSE) in a new dimensional context, venturing into the realm of non-Euclidean geometry. Our experimental validations on Printed Circuit Boards (PCBs) confirm the theoretical findings. Collectively, these results not only validate our theoretical framework but also demonstrate the potential of engineered circuit systems to emulate complex non-Hermitian phenomena, showcasing the applicability of non-Euclidean geometries in studying NHSE and topological phenomena in non-Hermitian systems.

In recent years, the study of non-Hermitian systems has attracted immense interest [1-12], primarily due to the non-Hermitian skin effect (NHSE) [13-24], a phenomenon where most eigenstates of non-Hermitian operators localize at the boundaries. This effect, characterized by edge or surface states that are adhered to the standard bulk-boundary correspondence (BBC), becomes particularly intriguing in its interplay with topological structures [15,17]. Topological invariants, such as the Zak phase and Chern number, derived from the Bloch Hamiltonian, are instrumental in identifying topological phases and phase transitions [25]. The experimental observation of NHSE in these contexts has led to a deeper understanding of topological phenomena in non-Hermitian systems [13,15,18,21].

Leveraging advanced electrical engineering techniques, electrical circuits have emerged as potent platforms for realizing and exploring topolectrical circuits [12,13,15,17,18,25-36]. These circuits, often comprising capacitors and inductors, have facilitated the realization of various topological states [25,26,28,31,32,34-36]. It is particularly evident in the domain of second-order topological insulators, where the topology is characterized by quantized quadrupole moment [32,34] and two-dimensional (2D) Wannier centers [35,36], as implemented in square and Kagome lattices. Notably, the incorporation of non-Hermitian and nonlinear characteristics through resistors, amplifiers, and nonlinear capacitors has expanded the scope of these experimental models. In 1854, the German mathematician Bernhard Riemann established a more comprehensive form of geometry, known as Riemannian geometry [37], of which Euclidean geometry is a special case. The establishment of Riemannian geometry provided the mathematical tools necessary for Albert Einstein to develop the General Theory of Relativity [38]. A quintessential example of Riemannian geometry is the geometry of a sphere. However, most experiments are still based on Euclidean geometric spaces [17,25-28,30,32-34,36] or employ simple passive components [17,25,26,28,32-34,36]. Only a limited number of experiments have utilized active components such as operational amplifiers [27,30], which leads to the pivotal question of our research: Can NHSE be realized and manipulated in more complex Riemannian geometric spaces?

In this work, we explore energy localization in non-Hermitian systems, focusing on the impact of topological defects in spherical models. Our investigation employs the Maximum Skin Corner Weight (MaxWSC) to analyze the influence of defects on the spatial distribution of modes. The work initiates with a defect-free non-Hermitian Su-Schrieffer-Heeger (SSH) model, examining NHSE through admittance spectra and inverse participation ratios (IPR) [39-41]. Subsequently, we introduce defects into the SSH chains to observe changes in mode distribution due to local perturbations. Inspired by recent findings on dimensional transmutation in non-Hermitian systems

[42], where non-Hermitian couplings alter the dimensionality of the system, we extend our research to Spherical Non-Hermitian Systems. In line with the principle that in 2D models, non-Hermitian couplings can lead to effective one-dimensional behavior, our spherical model considers a similar dimensional transition. By bisecting a sphere, we effectively create a system that can be conceptualized as one-dimensional chain-like structures, enabling us to study NHSE in a new dimensional context. This spherical model reveals diverse energy localization behaviors influenced by coupling strengths and defects. Theoretical findings are validated experimentally on Printed Circuit Boards (PCBs), confirming predictions of energy localization.

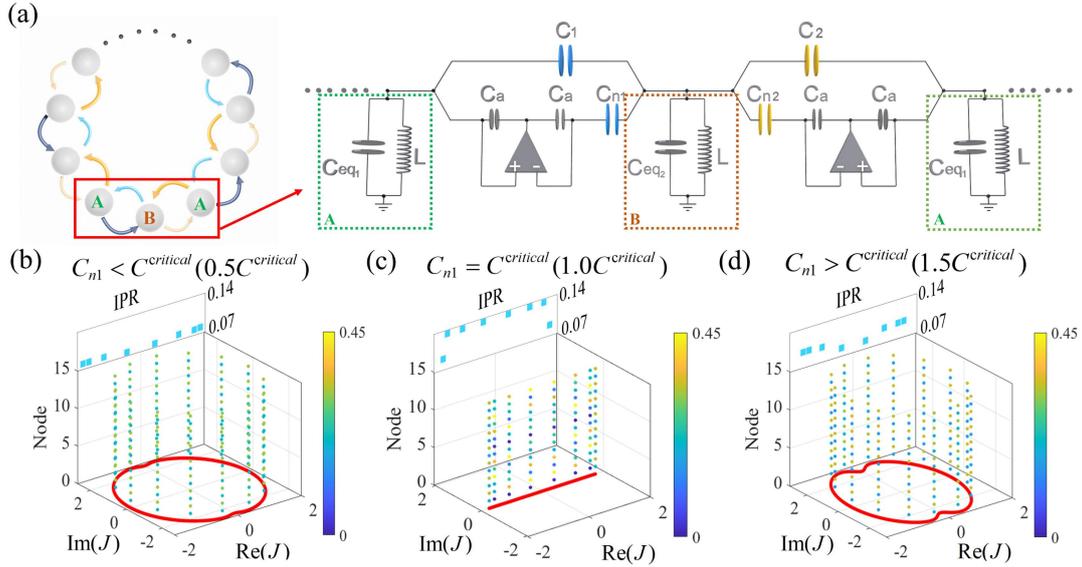

FIG. 1 Topological properties of defect-free cyclic non-Hermitian SSH chains. (a) The structural and circuit diagrams. (b-d) Admittance spectra under open boundary conditions indicated by red dots and periodic boundary conditions indicated by colored dots with defect-free cyclic structure. Inspection of the upper half of panels shows the IPR of eigenstates.

We initially consider a defect-free non-Hermitian SSH chain and its corresponding topolectrical circuits, as depicted in Fig. 1(a). Each unit cell comprises N=2 nodes, labeled as nodes A and B (enclosed within a red rectangular frame in Fig. 1(a)). The forward ($C_1 - C_{n1}$) and reverse ($C_1 + C_{n1}$) coupling between nodes A and

B, as well as the forward ($C_2 + C_{n2}$) and reverse ($C_2 - C_{n2}$) coupling between nodes B and A are distinct. The non-Hermitian characteristics of the SSH chain circuit are induced by $C_{n1}$ and $C_{n2}$. We achieve asymmetric coupling by connecting capacitors to unity-gain operational amplifiers. Notably, the sign of the non-Hermitian part of the coupling $C_{ni}$ can be flipped by inverting the operational amplifier, thereby reversing the bias between intercell and intracell segment couplings with respect to each other. This inversion of the sign of $C_{ni}$ in the asymmetric coupling, refer to as the negative impedance converter with current inversion (INIC) [27,30].

Following the Kirchhoff's laws, the Laplacian of the grounded circuit can be written as

$$L(\beta,\omega) = (j\omega)\begin{pmatrix} \frac{1}{\omega^2 L} - C_{eq_1} + (C_1 - C_{n1}) + (C_2 - C_{n2}) & (C_1 - C_{n1}) + (C_2 - C_{n2})\beta \\ (C_1 + C_{n1}) + (C_2 + C_{n2})\beta^{-1} & \frac{1}{\omega^2 L} - C_{eq_2} + (C_1 + C_{n1}) + (C_2 + C_{n2}) \end{pmatrix}. \quad (1)$$

Here, $\beta$ is the non-Bloch factor given by $\beta = re^{i\kappa} = e^{-\alpha}e^{i\kappa}$. The non-Bloch multiplication factor $\alpha$ denotes the exponential decay of the voltage and $\alpha$ is the non-Bloch multiplication factor. The diagonal terms of Eq. (1) represent a net shift in the eigenvalues of the Laplacian and can be adjusted by varying the frequency. At the resonant frequency of

$\omega = \frac{1}{\sqrt{LC_{xi}}}$ ( $C_{x1} = C_{eq_1} - (C_1 - C_{n1}) - (C_2 - C_{n2})$ $C_{x2} = C_{eq_2} - (C_2 + C_{n2}) - (C_2 + C_{n2})$ ),

the diagonal terms vanish.

Subsequently, we analyze the NHSE of the structure. As illustrated in Fig. 1(b-d), the red points on the coordinate plane represent the admittance spectra under periodic boundary conditions (PBC). The colored dots from blue to yellow in each column indicate the eigenmode at the complex admittance values given by the Re($J$) and Im($J$) under open boundary conditions (OBC). For each eigenmode, the $z$ coordinate of the dots indicates their spatial position along the chain, with the color of each point reflecting the energy value of that node.

To quantitatively assess the extent of localization within the eigenstates of our cyclic structure, we employ the IPR. The IPR for a given eigenstate $n$ is defined as [39-41]:

$$IPR_n = \sum_{i=1}^{L} \left|V_i^n\right|^4, \qquad (2)$$

where the summation extends over all lattice sites $i$ of the chain, and the eigenvectors $V_i^n$ are normalized such that $\sum_{i=1}^{L}\left|V_i^n\right|^4=1$. This normalization ensures that the IPR provides a consistent measure of localization across different states. Small IPR values are indicative of states that are extended or delocalized across the chain, whereas larger IPR values denote states that are highly localized. This metric is a crucial tool for distinguishing between extended and localized modes within the spectrum of our system.

Then, we delve into the criticality of non-Hermitian parameters in our system. The critical value of $C_{n1}$, pivotal in determining the extent of the NHSE, is intricately linked to the other non-Hermitian parameters and the coupling capacitors within our framework. Deriving from the complex admittance analysis, the formula for the critical value of $C_{n1}$ is given by [17]

$$C_{n1} = C^{critcal} = \frac{C_1 C_{n2}}{C_2}. \qquad (3)$$

This relationship suggests a delicate balance between these parameters, where any deviation from this critical value results in the localization of voltage eigenstates, a hallmark of NHSE. Consequently, understanding and manipulating $C_{n1}$ within its critical range becomes crucial for controlling and exploring the NHSE in our system.

The calculations for $C_{n1} < C^{critical}$, $C_{n1} = C^{critical}$, $C_{n1} > C^{critical}$ have been conducted, with the same values of $C_1 = 1$ $\mu F$, $C_2 = 1$ $\mu F$ and $C_{n2} = 0.5$ $\mu F$. $C_{n1}$ is set to $0.5\ C^{critical}$, $1.0\ C^{critical}$ and $1.5\ C^{critical}$ in Fig. 1(b), (c) and (d), respectively. For these three situations, it is observed that the eigenmode at each complex admittance value spatially coincides with the admittance spectra under PBC conditions. Inspection of the upper half of panels in Fig. 1(b-d) reveals that the calculated IPR values do not exhibit significant magnitudes across the three cases. The absence of elevated IPR values suggests that the extended states within our system do not gradually localize with nonzero.

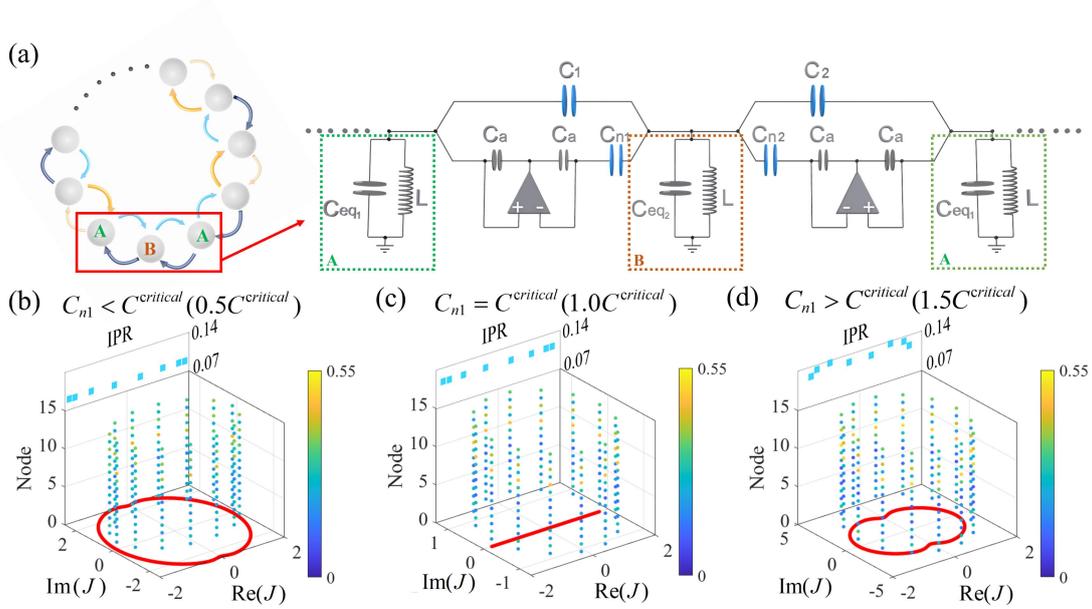

FIG. 2 Topological properties of defective cyclic non-Hermitian SSH chains. (a) The structural and circuit diagrams. (b-d) Admittance spectra under OBC indicated by red dots and PBC indicated by colored dots with defective cyclic structure. Inspection of the upper half of panels shows the IPR of eigenstates.

Subsequently, we explore the incorporation of defective states into an originally pristine cyclic structure. We now modify the coupling at one specific site in the cyclic structure to match the coupling at adjacent sites as shown in Fig. 2(a). The defect point is set at node 8. This modification entails changing the original forward coupling ( $C_2 + C_{n2}$ ) to ( $C_1 - C_{n1}$ ) and the reverse coupling ( $C_2 - C_{n2}$ ) to ( $C_1 + C_{n1}$ ).

It is observed that the spatial distribution of the eigenmode at each complex admittance value does not coincide with the admittance spectra under PBC in Fig. 2(b-d). The deviation highlights the impact of introducing non-Hermitian defect states in the system. The altered coupling due to the defects leads to a distinct spatial distribution of modes, illustrating the sensitivity of the system's spectral characteristics to local perturbations. Upon examination of the IPR values presented in the upper half of panels in Fig. 2(b-d), we observe a same across the three considered cases. For the three cases, the IPR values remain very low, indicating that the extended states retain their delocalized character without significant signs of localization. We can see from the graph that energy mainly concentrates near the defect location.

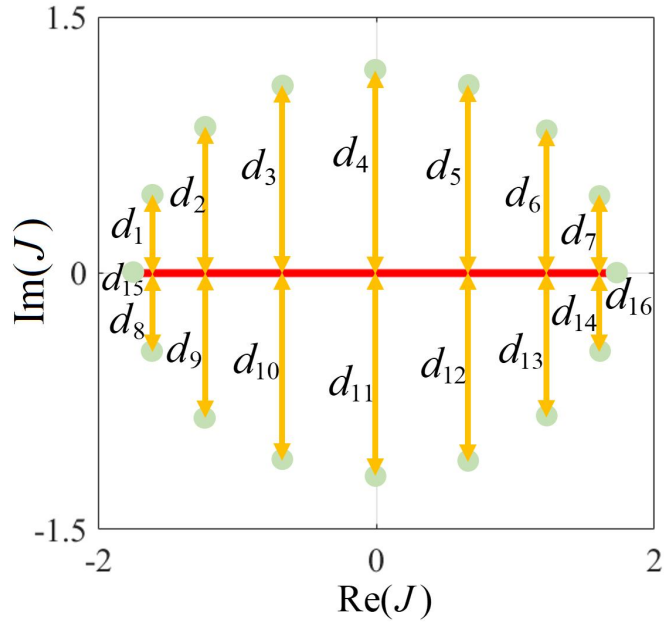

FIG. 3 Quantitative analysis of the admittance spectra deviation.

To quantify the extent to which the eigenmode of various structures approximate the PBC admittance spectra, we introduce the following formula:

$$D = \sum_{i=1}^{n} d_i = \sum_{i=1}^{n} \left\{ \sqrt{\left[\text{Re}(\lambda_i) - \text{Re}(\lambda_{PBC_i})\right]^2 + \left[\text{Im}(\lambda_i) - \text{Im}(\lambda_{PBC_i})\right]^2} \right\}. \quad (4)$$

An instantiation of Eq. (4) is presented in Fig. 3, which showcases a top view of the eigenmode distributions in relation to the PBC admittance spectrum, incorporating a structural defect. In this specific situation, we have designated the coupling coefficient $C_{n1} = C^{critical}$ and the defect is introduced at node 7. We define $d_i$ as the distance of each OBC eigenmode admittance spectra from the PBC admittance spectra. The sum of $d_i$ across all characteristic modes is computed, yielding the results as per Eq. (4).

TABLE. 1 $D$ values for defect-free cyclic structure and defective cyclic structure.

| Node | cyclic structure (defect-free) | cyclic structure (defective) |
|---|---|---|
| 8  | $1.2045 \times 10^{-2}$ | 11.848 |
| 10 | $1.0317 \times 10^{-2}$ | 11.946 |
| 12 | $1.3561 \times 10^{-2}$ | 11.998 |
| 14 | $1.3296 \times 10^{-2}$ | 12.029 |
| 16 | $1.5160 \times 10^{-2}$ | 12.049 |

Subsequent calculations of the $D$ values for the cyclic structure, both in the presence and absence of defects at various nodes, are executed by Eq. (4) and delineated in Table 1. The data reveals that the $D$ values for structures without defects are three orders of magnitude lower than those with defects. Within a certain margin of computational error, it supports the conclusion that "the eigenmode at each complex admittance value spatially coincides with the admittance spectra under PBC conditions". The markedly lower $D$ values in defect-free structures suggest a strong alignment with PBC, indicating that the defect free circular structure belongs to PBC.

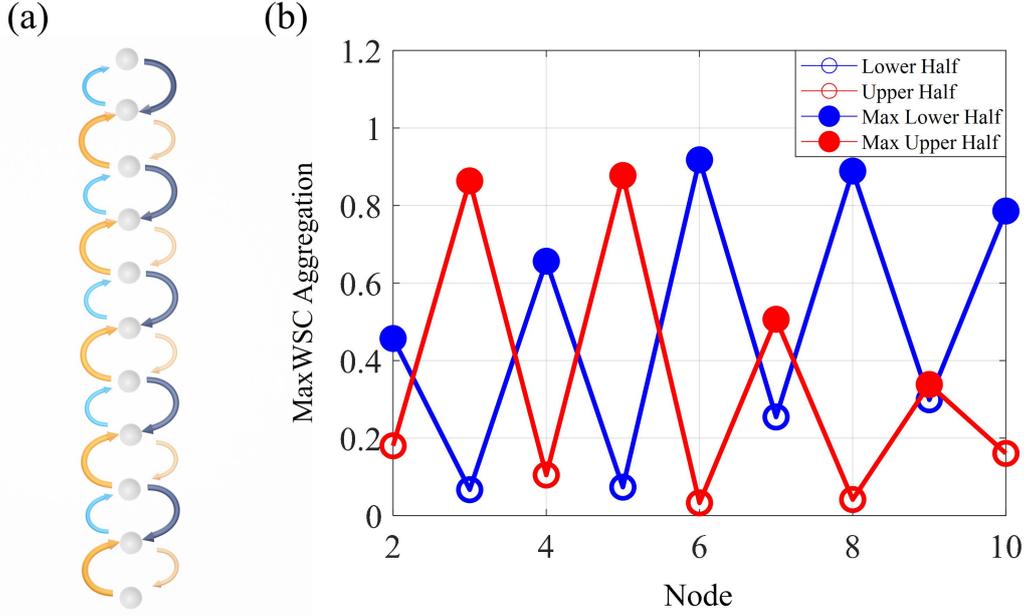

FIG. 4 (a) Schematic representation for a linear chain lattice. (b) MaxWSC as a function of node defects position, quantifying the energy aggregation in the system. The blue line represents the MaxWSC for the lower half of the chain, while the red line corresponds to the upper half. Solid circles indicate the MaxWSC values for each half, highlighting regions of peak energy concentration.

To reveal the longitudinal characteristics in spherical systems, we take care of the behavior of an open chain with N=11 nodes depicted in Fig. 4. The analysis presented in Fig. 4(b) focuses on the MaxWSC as a function of node position, providing a quantification of energy concentration within the system. We define MaxWSC(i) as

$$\text{MaxWSC(Node)} = \max_{1 \leq i \leq n}[\sum_j |V(j,i)|^4 e^{(-|r(j)-r(Node)|/\varepsilon)}], \qquad (5)$$

where $V(j,i)$ represents the $j$-th eigenvector corresponding to the $i$-th eigenvalue. The first factor $|V(j,i)|^4$ measures the localization strength of eigenmodes. The second factor $e^{(-|r(j)-r(Node)|/\varepsilon)}$, which decays exponentially away from the corner with decay length $\varepsilon$, selects only the modes localized at the corners. MaxWSC (Node) represents the maximum weighted state contribution at a given node within the system. The summation of MaxWSC values for each half provides a clear indication of energy localization: a higher cumulative MaxWSC in one half implies a greater concentration of energy in that region. By systematically analyzing these sums for different defect configurations, we can know the role that the position of a defect plays in directing the energy distribution pattern within the model.

Then we examine the energy distribution across a network structure described by an $n \times n$ Laplacian matrix. Each eigenmode of this matrix corresponds to a distinct energy state of the system, with $n$ eigenvectors delineating the state's distribution across the nodes of the network. To quantify the contribution of each state to one given node, we calculate the WSC for every eigenmode relative to the node, utilizing the IPR. The IPR serves as a localization metric, indicating the extent to which a state is localized around a particular node. For one given node, we determine the MaxWSC by identifying the maximum WSC value from among all the eigenvalues, thereby gauging the peak energy presence at that node. Subsequently, by computing the MaxWSC for each node sequentially, we construct a profile of energy distribution for the entire structure shown in FIG. S1. It allows us to visualize and analyze the pattern of energy localization, which is particularly influenced by the network's topological defects. Regions with higher MaxWSC values indicate nodes with greater energy localization, which are critical for understanding the flow and distribution of energy within the network, as well as the influence of non-Hermitian properties.

The term "Node=2" in Fig. 4(b) refers to the introduce of defects characterized by modifying the coupling strength between the second and third nodes equal to that between the first and second nodes, which highlights the influence of symmetric coupling variations within the system's structure. Our observations point to a correlation between the position of the defects and the localization of energy: when the defect is situated at an odd-numbered node, the energy tends to be localized in the lower half of the chain; conversely, when the defect is at an even-numbered node, energy localization is observed in the upper half. We also provide the MaxWSC of nodes with defective between layers in Fig. S2.

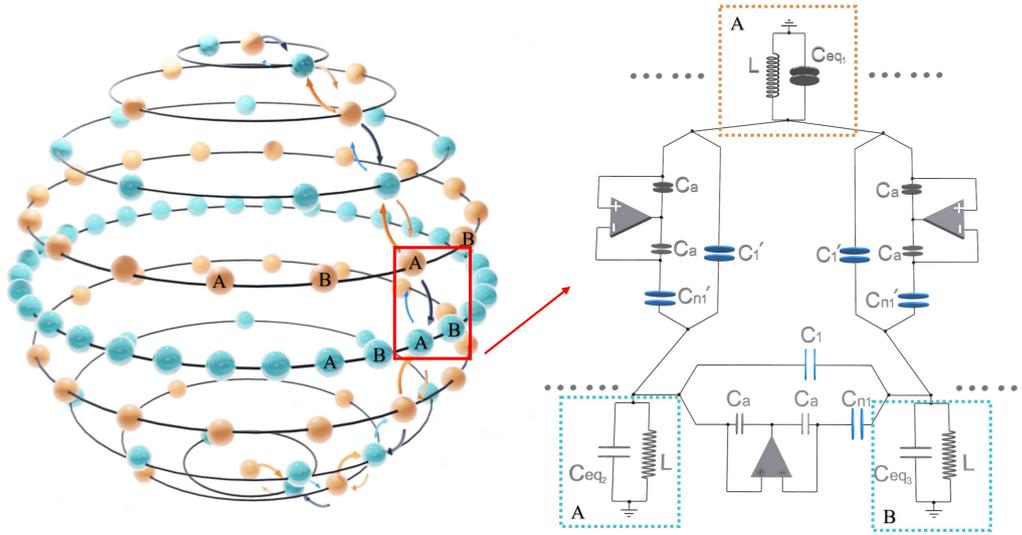

Fig. 5 Structure diagram and partial circuit diagram of the spherical model.

Non-Hermitian couplings can lead to a dimensional reduction in the effective Brillouin zone of lattices, a phenomenon termed "dimensional transmutation" [42]. It challenges traditional classifications of topological states by showing that topological invariants in non-Hermitian systems can be protected by lower-dimensional topologies. By bisecting the spherical model along the equator and unfolding it into a plane, topological properties are preserved even through the mapping from non-Euclidean geometry to Euclidean geometry space, aligning with findings on dimensional transmutation in non-Hermitian systems without altering topological characteristics.

The spherical model is comprised of 11 concentric layers that mimic the stratification observed in spherical coordinates. These layers are populated with nodes that increase and then decrease in number from the top to the bottom, specifically in the sequence of 1, 2, 4, 8, 16, 32, 16, 8, 4, 2, and 1, respectively. Each layer constitutes a closed ring circuit, analogous to the cyclic model previously discussed. The interlayer couplings are designed to emulate the chain-like structure, resulting in an asymmetric coupling between consecutive layers. For this structure, there are 94 nodes in total, and we specify the top node number to 1.

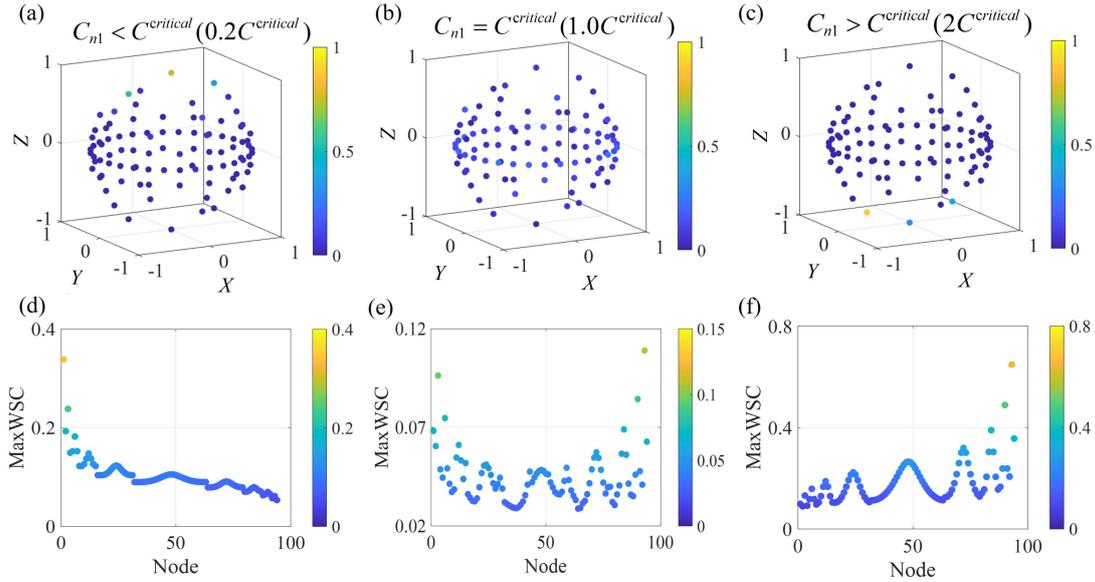

FIG. 6 Admittance spectra and energy localization in the spherical model under different conditions. (a-c) Illustrate the admittance spectra of the spherical model for three distinct cases at a specific eigenvalue. (d-f) Display the MaxWSC for the corresponding spherical model cases, with the node index plotted along the horizontal axis and the MaxWSC value along the vertical axis.

The ramifications of varying inter-node coupling strengths within the spherical circuit model are delineated in Fig. 6, each possessing an inherent critical coupling threshold—$C_1^{critical}$ for intra-ring and $C_2^{critical}$ for inter-ring interactions. To facilitate a streamlined discussion of the system's behavior under various regimes, these critical couplings are equated in the analysis. The circuit parameters are meticulously chosen as follows: $C_1 = C_1^{'} = 1.8\ \mu F$, $C_2 = C_2^{'} = 1\ \mu F$ and $C_{n2} = C_{n2}^{'} = 0.5\ \mu F$. This parametric configuration allows us to probe into three distinct regimes characterized by the relative strength of $C_{n1} < C^{critical}$, $C_{n1} = C^{critical}$, and $C_{n1} > C^{critical}$.

Energy distributions corresponding to discrete eigenvalues are rendered into a spherical coordinate system in Fig. 6(a-c), visualized in a three-dimensional representation with the *X, Y,* and *Z* axes corresponding to the physical structure. The color bar indicates the magnitude of the admittance at each node. The distribution shown in Fig. 6(a) is characterized by a predominant allocation of energy to the upper hemisphere of the sphere. Contritely, a state of minimal energy aggregation, with no discernible localization within the spherical domain, is represented in Fig. 6(b). Conversely, portrayed in Fig. 6(c) is a concentrated energy disposition within the lower hemisphere, indicating an inverse localization pattern relative to that observed in Fig. 6(a). The MaxWSC computations, illustrated in Fig. 6(d-f), provide a quantitative synthesis of the localization phenomena observed. A predilection for energy concentration at the top of the sphere is displayed in Fig. 6(d), whereas a gravitation of energy towards the bottom is revealed in Fig. 6(f). The intermediate scenario, captured in Fig. 6(e), evidences a more homogeneous energy distribution across the polar extremes of the sphere, with MaxWSC magnitudes that are diminutive in comparison to the other examined cases, hence indicating a less conspicuous congregation of energy. In addition, we provide MaxWSC Aggregation of node defects in spherical model with three cases in Appendix B.

Thus, the spherical model, an intricate amalgamation of ring and chain motifs subject to different coupling strengths, exhibits a rich tapestry of energy localization behaviors that are contingent upon the critical coupling constants. By calibrating $C_1$ and $C_2$ to their critical counterparts and systematically adjusting $C_{n1}$, we have unveiled distinct localization regimes. These regimes are characterized by their energy distributions, which transition from polar concentration to more uniform dispersion across the spherical lattice.

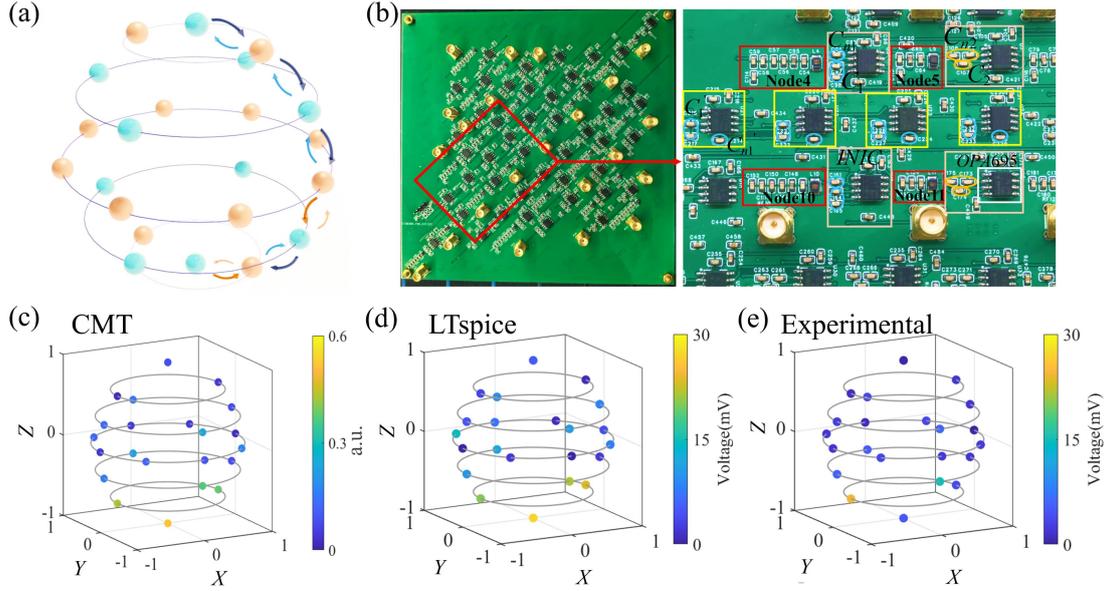

FIG. 7 Experimental realization and energy distribution mapping of the spherical model on a PCB. (a) The simplified schematic of the seven-layered spherical model with an intentional defect introduced at Layer 2. (b) The photograph of the PCB fabricated to embody the spherical model. Insets provide a magnified view of critical nodes, illustrating the circuitry and components that constitute the system. (c-e) The admittance spectra under CMT, LTspice simulation, and experimental.

A comprehensive portrayal of the empirical validation for energy distribution in the presence of inter-layer defects within a spherical model, achieved through the construction and analysis of a seven-layer physical circuit, is provided in Fig. 7. The overarching PCB layout of the assembled circuit, along with a detailed view of the critical components, is displayed in Fig. 7(b). The global layout (left) and the zoomed-in sections (right) are annotated to identify the nodes (red boxes), the coupling circuits between the ring structures (light pink boxes), and the inter-layer coupling circuits (yellow boxes). The INIC within our experimental setup is composed of OPA695 operational amplifiers, chosen for their disabled function and ultra-wideband current feedback capabilities, provided by Texas Instruments.

The circuit parameters are meticulously calibrated to $C_1 = C_1' = 2\ \mu F$, $C_2 = C_2' = 1\ \mu F$, $C_{n2} = C_{n2}' = 0.5\ \mu F$ and $L = 10\ mH$. Defects were systematically introduced at Layer 2 under the condition $C_{n1} = C^{critical}$. The admittance spectrum shown in Fig. 7(c) corroborates the theoretical prediction by

CMT, with energy prominently localized in the lower hemisphere, consistent with the results presented in Fig. S2 that defects in even-numbered layers behave in such a distribution.

To reinforce these findings, a direct circuit simulation by LTspice is conducted, mirroring the physical circuitry to ascertain the voltage values at each node at the circuit's resonant frequency of 15.91 kHz. The resultant energy distribution from the simulation, depicted in Fig. 7(d), aligns with the admittance spectrum, confirming the lower hemisphere energy concentration. The experimental verification, as shown in Fig. 7(e), also indicates energy localization in the lower hemisphere. Slight discrepancies between the measured node voltages and the simulation data can be attributed to the inherent tolerance in the circuit components—inductors with a 2% margin of error and capacitors with a 5% margin, as well as the precision limits of the measurement apparatus. Despite these variances, the overall experimental outcomes resonate with the theoretical model, signifying a successful experimental observation of the predicted energy localization effects within the meticulously designed circuit. Details of the specific measurement instruments can be found in Appendix C.

In summary, we investigate energy localization in non-Hermitian systems, focusing on the effects of topological defects in cyclic and spherical models. Employing the MaxWSC, we begin with a defect-free non-Hermitian SSH model, studying NHSE through admittance spectra and IPR. We observe altered mode distributions Introducing defects into the SSH chains, then our research progresses to spheres geometries, inspired by the concept of dimensional transmutation where non-Hermitian couplings alter system dimensionality [42]. Specifically, we extend this idea to our spherical model, bisecting a sphere to create chain-like structures for NHSE. This model unveils diverse energy localization behaviors influenced by coupling strengths and defects. Our theoretical findings are confirmed experimentally on a PCB. This study not only extends the experimental exploration of non-symmetric coupling models to Riemannian geometric spaces but also enhances their applications in microscale energy manipulation and control within electronic circuits.

**Acknowledgements**

The authors thank for NUPTSF (Grants No. NY220119, NY221055).

# APPENDIX A: SUPPLEMENT TO MaxWSC Aggregation

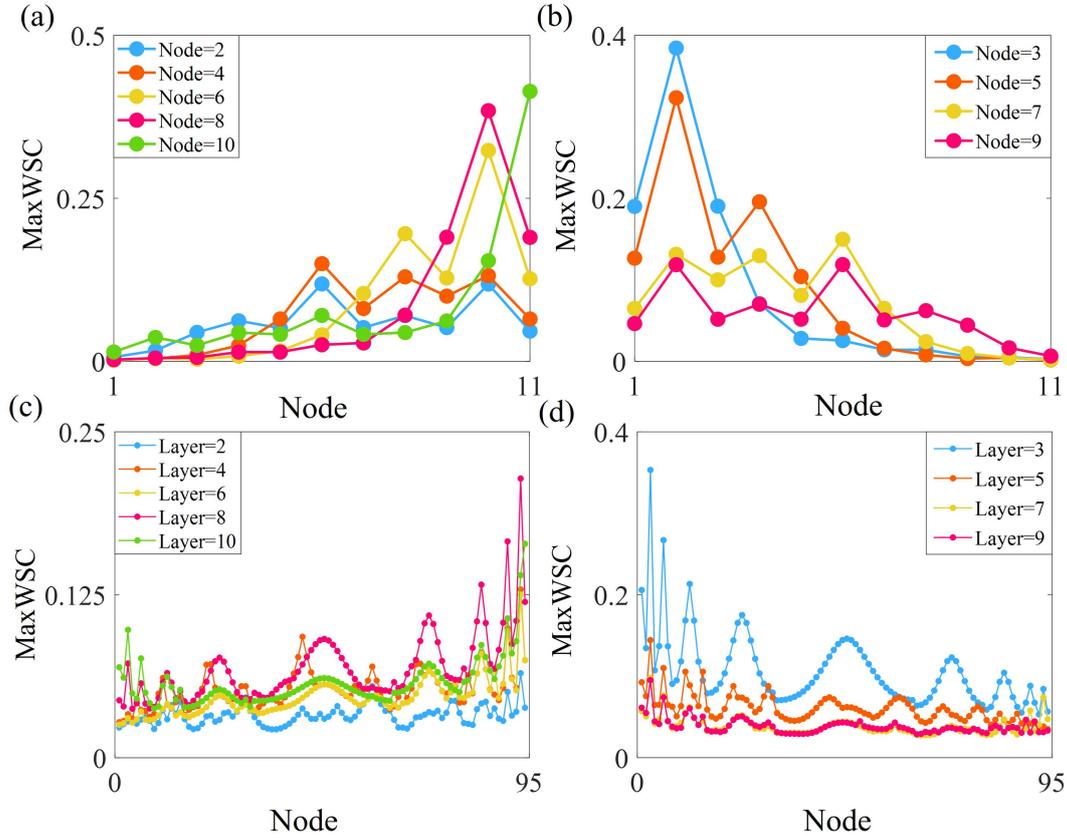

FIG. S1 Energy analysis of defect states in chain and spherical model using MaxWSC. (a) MaxWSC for defects at even nodes in the chain model. (b) MaxWSC for defects at odd nodes in the chain model. (c) MaxWSC for defects in even layers of the spherical model. (d) MaxWSC for defects in odd layers of the spherical model.

In the analysis shown in Fig. S1, which extends the observations from Fig. 4 and Fig. 7, the focus is on examining the MaxWSC values across various defect states. The data in S1. (a-b) reveals that when defects are positioned at even-numbered nodes, there is a notable concentration of MaxWSC at the lower nodes. On the other hand, defects at odd-numbered nodes result in an accumulation of MaxWSC at the upper nodes. This distribution demonstrates consistency with the patterns and conclusions in Fig. 4.

The results provide focus on the MaxWSC distribution in spherical model with layered defects in Fig. S1(c-d). This examination specifically scrutinizes how MaxWSC varies across different layers within the spherical model. It becomes evident that defects in even layers lead to MaxWSC predominantly localizing in the sphere's lower hemisphere. In contrast, defects in odd layers are associated with MaxWSC primarily concentrating in the upper hemisphere. This distinct pattern of MaxWSC distribution not only complements but also substantiates the findings observed in Fig. 7, suggesting a robust interplay between the positioning of defects and MaxWSC localization in these complex systems.

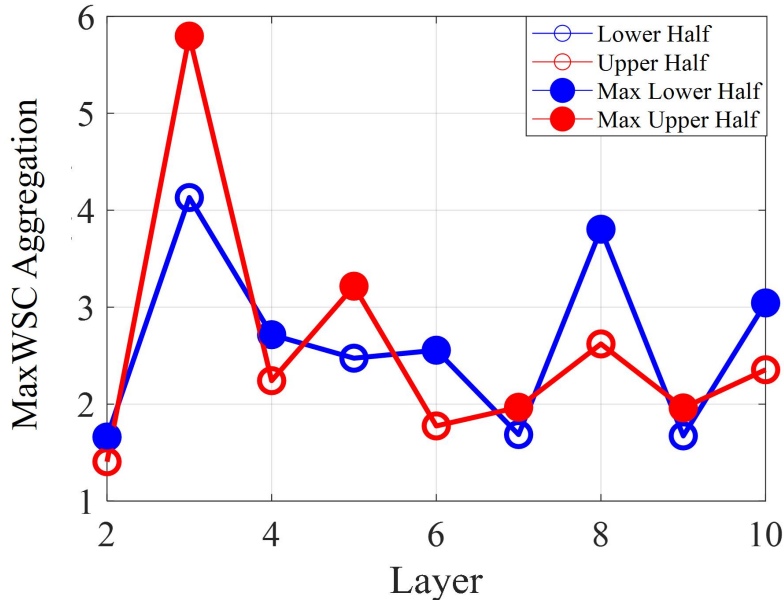

FIG. S2 The MaxWSC as a function of the layer position within the spherical model, illustrating the energy aggregation in response to interlayer defects. The blue line with open circles denotes the MaxWSC for the lower hemisphere of the sphere, while the red line with open circles represents the MaxWSC for the upper hemisphere. The solid circles highlight the peak MaxWSC values for each hemisphere, identifying the layers with the most pronounced energy concentration.

In delineating the energy landscape of our spherical model, 'Layer=2' is designated to represent a configuration where the coupling strength between the first- and second-layers mirrors that of the coupling between the second and third layers. We set the inter-layer coupling parameter $C_{n1}$ at parity with the system's critical coupling $C^{critical}$. To discern the influence of localized structural perturbations, we introduce defects sequentially to each layer and performed a comprehensive sweep to evaluate the MaxWSC under these varying defect conditions. The graphical representation of this analysis is presented in Fig. S2, which plots the MaxWSC against the layer index, offering a quantitative measure of energy localization in the presence of layer-specific defects.

The emergent pattern, as depicted in Fig. S2, aligns with observations from our chain-like structure shown in Fig. 4(b), where defects in odd-numbered layers lead to a preponderance of energy in the lower hemisphere, and conversely, defects in even-numbered layers result in the accumulation of energy in the upper hemisphere. This consistent trend not only draws parallels between the spherical model's inter-layer coupling and that observed in chain-like structures but also underscores the analogous influence of structural asymmetries on energy distribution. The symmetry of the energy distribution in response to defects reveals underlying principles of non-Hermitian physics, where the spatial configuration plays a pivotal role in governing the localization dynamics.

# APPENDIX B: SUPPLEMENT OF NODE DEFECTS UNDER SPHERICAL MODEL

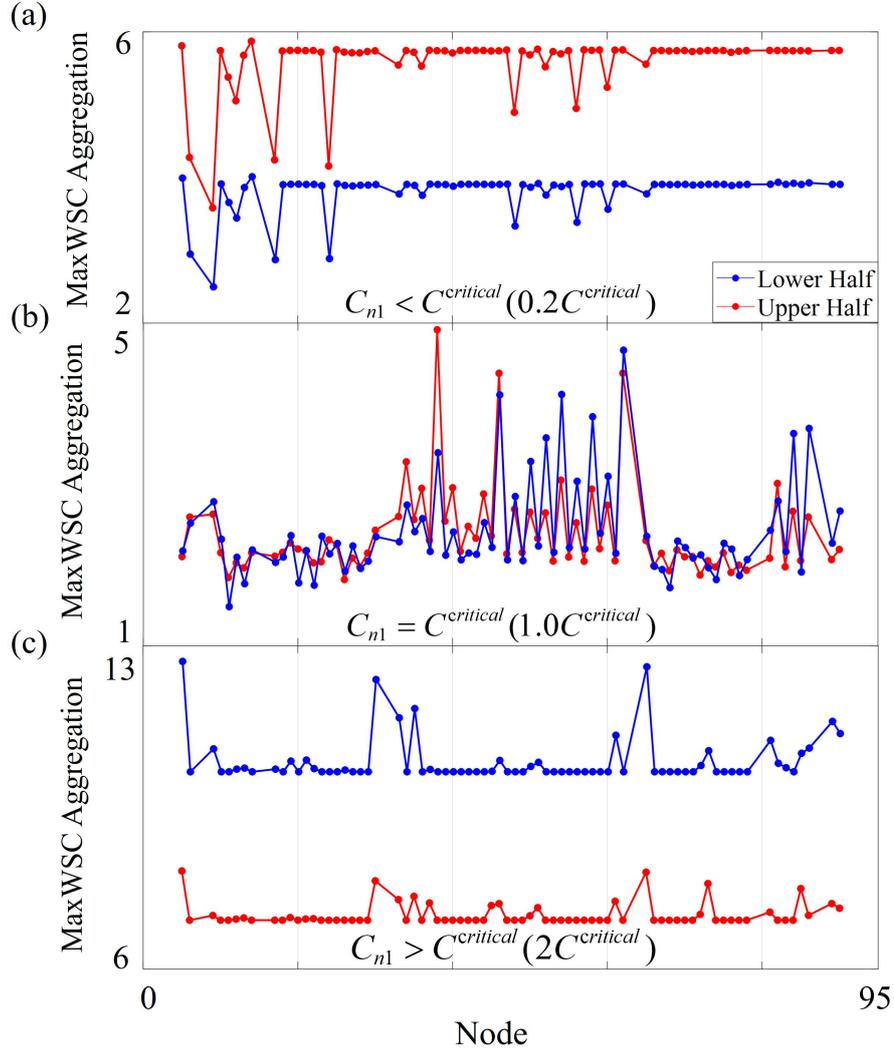

FIG. S3 MaxWSC Aggregation of node defects in spherical model.

We conduct a study on energy localization at node defects in cyclic model, observing a predominant congregation of energy at these defect sites in Fig. 2. We extend this analysis to spherical model, where we examine the aggregation of energy at every node defect. Our discussion is structured around three distinct scenarios: $C_{n1} < C^{critical}$, $C_{n1} = C^{critical}$, $C_{n1} > C^{critical}$. The energy accumulation with defects added in Fig. S3 (a) and (c) is consistent with that without defects in Fig. 6. The MaxWSC Aggregation values in the upper and lower parts of Fig. S3 (b) do not remain separate. Indicates that energy accumulation is not as obvious.

# APPENDIX C: CIRCUIT MEASUREMENT AND TESTING INSTRUMENT

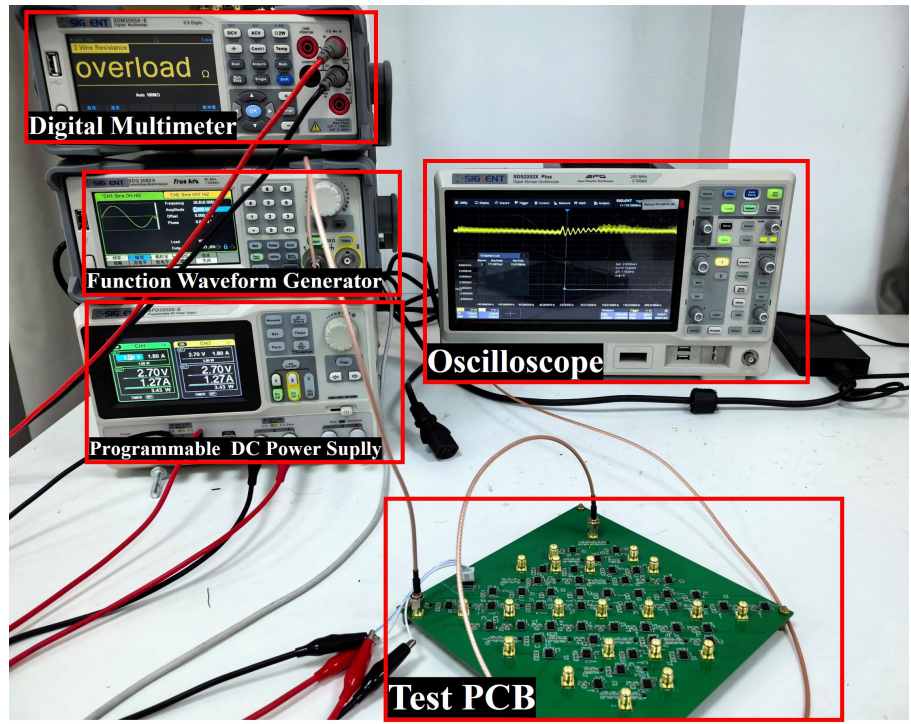

FIG. S4 Photos of PCB and testing instruments.

We utilize JLC EDA software to develop PCB, and make appropriate designs for PCB composition, stacking layout, inner layer, and grounding design. To avoid signal crosstalk during experimental testing. We use SMA interfaces and coaxial cables to connect the oscilloscope and signal generator. Our laboratory uses a complete set of SIGLENT instruments for measurement. The Oscilloscope is SDS2202X Plus, the Function Waveform Generator is SDG 2082X, the Programmable DC Power Supply uses SPD3303X-E, and the Digital Multimeter uses SDM3055X-E.